\documentclass[12pt,a4paper]{article}
\usepackage{amsmath}
\usepackage{graphicx}
\usepackage{ulem}
\usepackage{xcolor}
\usepackage{times}
\usepackage{cite}
\begin{document}
	\begin{titlepage}
		\title{Reflective ability and its energy dependence}
		\author{ S.M. Troshin, N.E. Tyurin\\[1ex]
			\small   NRC ``Kurchatov Institute''-- IHEP\\
			\small   Protvino, 142281, Russian Federation}
		\normalsize
		\date{}
		\maketitle

\begin{abstract}
	We introduce the notion of  reflective ability and discuss its energy dependence	using  rational form of unitarization.  Correspondence with the phase of exponential unitarization is traced. Increase of the reflective ability of  interaction region  starts at the LHC  energies. Numerical estimates  are given.

\end{abstract}

\end{titlepage}
\section{Introduction. Reflective ability}

Introduction of the reflective ability notion aims to clarify and provide another test for asymptotics  under  the hadron interactions. The studies of dynamics  of the elastic hadron scattering is considered to be a relevant tool for achieving that purpose.

The elastic scattering element of $S$--matrix  in the impact parameter space denoted in what follows as $S$ ($S\equiv S(s,b)$) is the real function in case of a pure imaginary approximation for the elastic scattering amplitude.   So, this function can  be either positive or negative. 
Nonnegative values of $S$ correspond to the absorptive scattering while the negative ones were interpreted as a result of the reflective scattering \cite{07} by analogy with optics \cite{land}. 
The reflective scattering mode can be associated with  central core presence in a hadron structure. 

Unitarity provides the  constraint  $|S|\leq 1$. 
It is quite natural to introduce the reflective ability, when $S<0$, as 
\begin{equation}
	R(s)\equiv|S(s,0)|.
\end{equation}
This definition corresponds to the  absorptive ability definition when $S\geq 0$. Reflective ability is due to the central geometric elastic scattering while the absorptive ability is about peripheral shadow elastic scattering \cite{cent}. 

We consider here   the reflective ability $R(s)$  on base of rational unitarization and demonstrate that its increase is ia most probable  behavior.  \textcolor{blue}{It should be noted that there are  interesting results obtained in perturbative QCD  in the framework of parton model (see \cite{papa} and references therein). Those provide the results for the asymptotic regime of soft scattering summing up leading terms of $\ln s$ in  each order of perturbative QCD expansion. The observed states are colorless hadronic states and  confinement  can modify predictions. In its turn unitarity being formulated for  asymptotic hadronic states  could give an important insight on the confinement problem.} 
\section{Rational unitarization vs exponential one}
Unitarization of an ``input amplitude'' is a commonly used approach to obtain final  amplitude consistent with  unitarity. The recent discussion of the unitarization method and choice of the input for this procedure has beed given in \cite{sym}. Indeed, unitarization is  mapping of  some input  quantity to the unitary circle. We discuss the two kinds of mapping,  the rational and exponential ones. There are also  hybrid approaches which we will not concern here.

Under the rational unitarization
the function $S(s,b)$ is expressed through the "input amplitude" $U(s,b)$ by the following ratio:
\begin{equation}\label{map}
	S(s,b)=\frac{1+iU(s,b)}{1-iU(s,b)},
\end{equation}
which is a one-to-one transform between the functions $S$ and $U$. Eq. (\ref{map}) performs mapping  upper half-plane of complex domain of allowed by unitarity $U$-variation onto the unit circle of explicit unitary $S$--variation domain (Fig. 1).
\begin{figure}[hbt]
	\hspace{-1cm}
	\vspace{-2cm}
	\resizebox{14cm}{!}{\includegraphics*{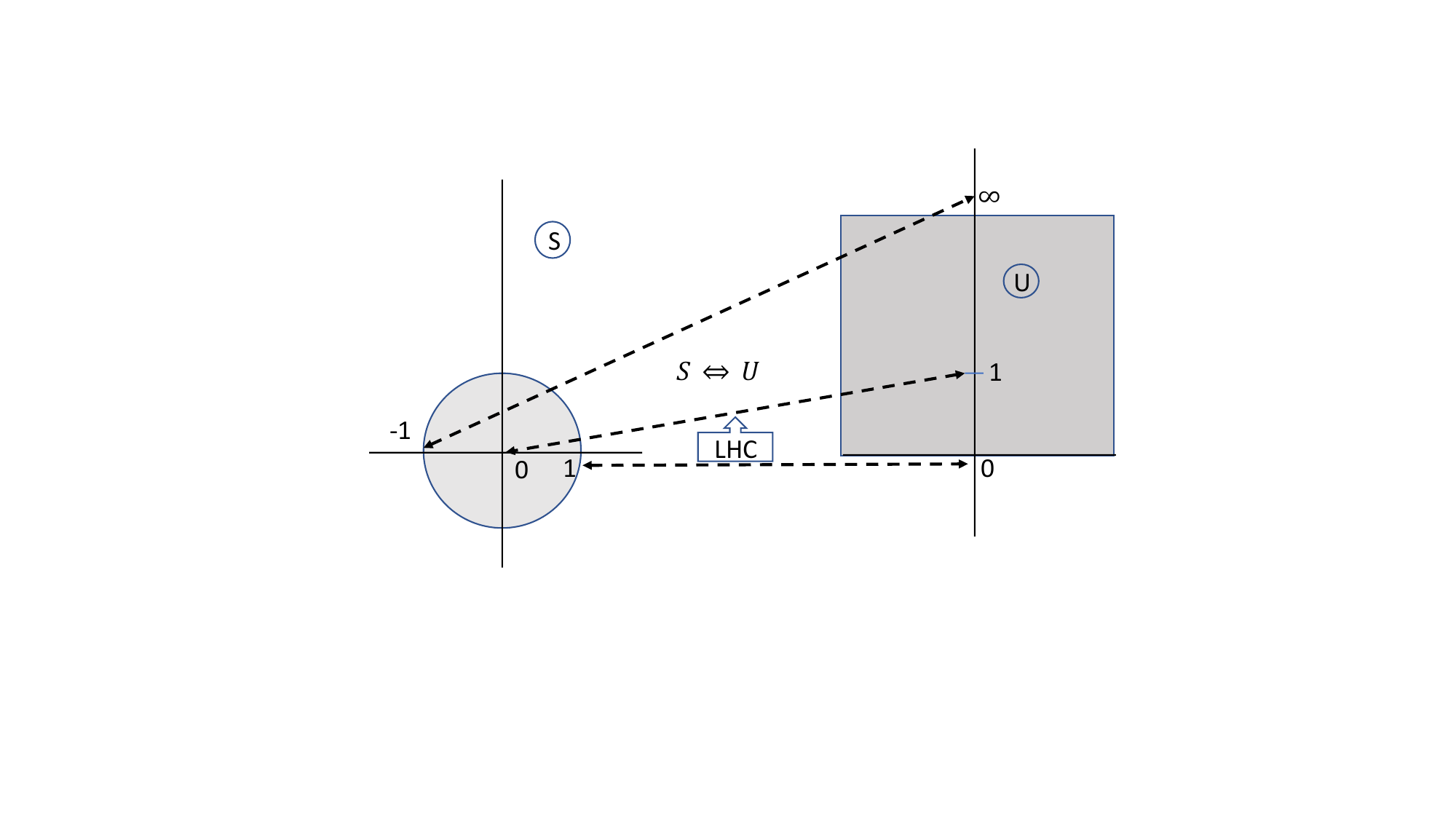}}
	
	\caption[ch2]{\small One-to-one  mapping by the rational unitarization, Eq. (\ref{map}). The LHC energies correspond to $S\simeq 0$.}
\end{figure}
In mathematics, it is known as Cayley transform, the particular case of Mobius transform. 

It should be noted that the exponential unitarization
\begin{equation}\label{exp}
	S(s,b)=\exp[2i\delta(s,b)]
\end{equation}
with $\delta(s,b)\equiv \delta_R(s,b)+i\delta_I(s,b)$ is not a one-to-one transform. It corresponds to $S\neq 0$ for any finite value of $\delta$ (Fig. 2).
\begin{figure}[hbt]
	\hspace{-1cm}
	\vspace{-2cm}
	\resizebox{14cm}{!}{\includegraphics*{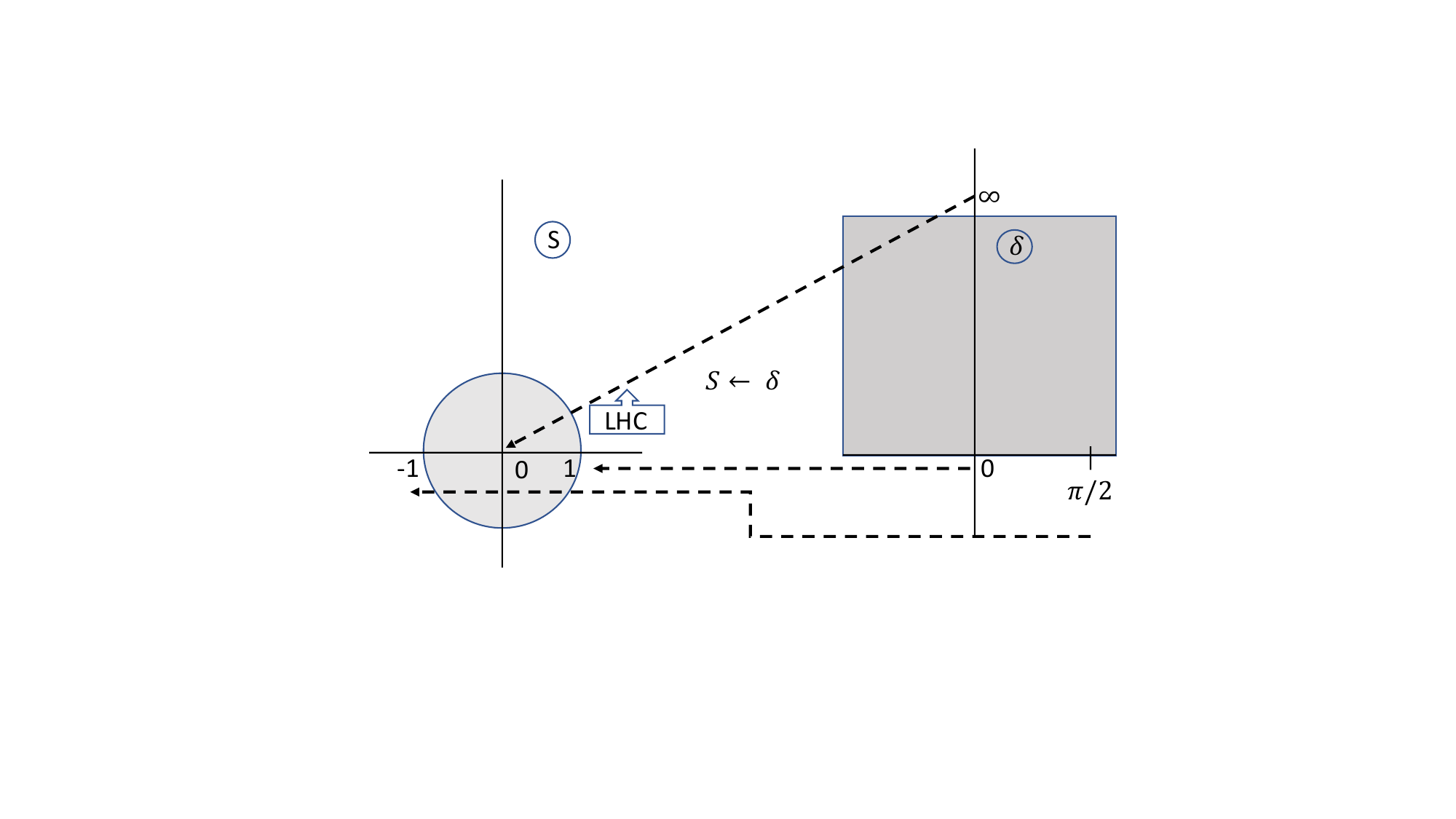}}
	
	\caption[ch2]{\small One--way mapping by the exponential unitarization,  Eq. (\ref{exp}). }
\end{figure}
The both  mappings are conformal and due to unitarity Im$U\geq 0$ as well as Im$\delta\geq 0$. Expression 
for the phase $\delta(s,b)$ through the function $U(s,b)$ and the respective phase features have been discussed in \cite{07}.

Thus, we consider the rational unitarization since it is simple and one-to-one transform, provides a smooth transition to the reflective scattering mode and is consistent with physically motivated  damping of radiation \cite{gol}. The corresponding phase behavior is also traced.  

We refer to
pure imaginary case of the scattering, i.e. under replacement $U\to iU$ we rewrite Eq. (\ref{map}) in the simplified form 
\begin{equation}\label{mapi}
	S(s,b)=\frac{1-U(s,b)}{1+U(s,b)}
\end{equation}
The basic principles  of the $U$--matrix construction have recently been discussed in \cite{sym}.
The arguments  based on the geometrical models of hadron collisions were also given. 
The factorized form 
\begin{equation}\label{usb}
	U(s,b)=g(s)\omega(b),
\end{equation}
with the dependence $g(s)\sim s^\lambda$ was proposed. 
The function $\omega(b)$ can be interpreted as a convolution of the two hadron matter distributions in the transverse plane with the dependence $\omega(b)=\exp(-\mu b)$, $\mu>0$ in the geometrical models of hadron collisions. 
\section{Energy dependence of  reflective ability}
The reflective scattering ($S<0$) appears first at $b=0$, i.e. when $U(s,0)>1$ in Eq. (\ref{mapi})  at $s>s_r$ under energy increase and the reflective ability $R(s)$ has the form
\begin{equation}\label{mapr}
	R(s)=[g(s)-1]/[g(s)+1].
\end{equation}
\begin{figure}[hbt]
	\hspace{0cm}
	\vspace{-2cm}
	\resizebox{16cm}{!}{\includegraphics*{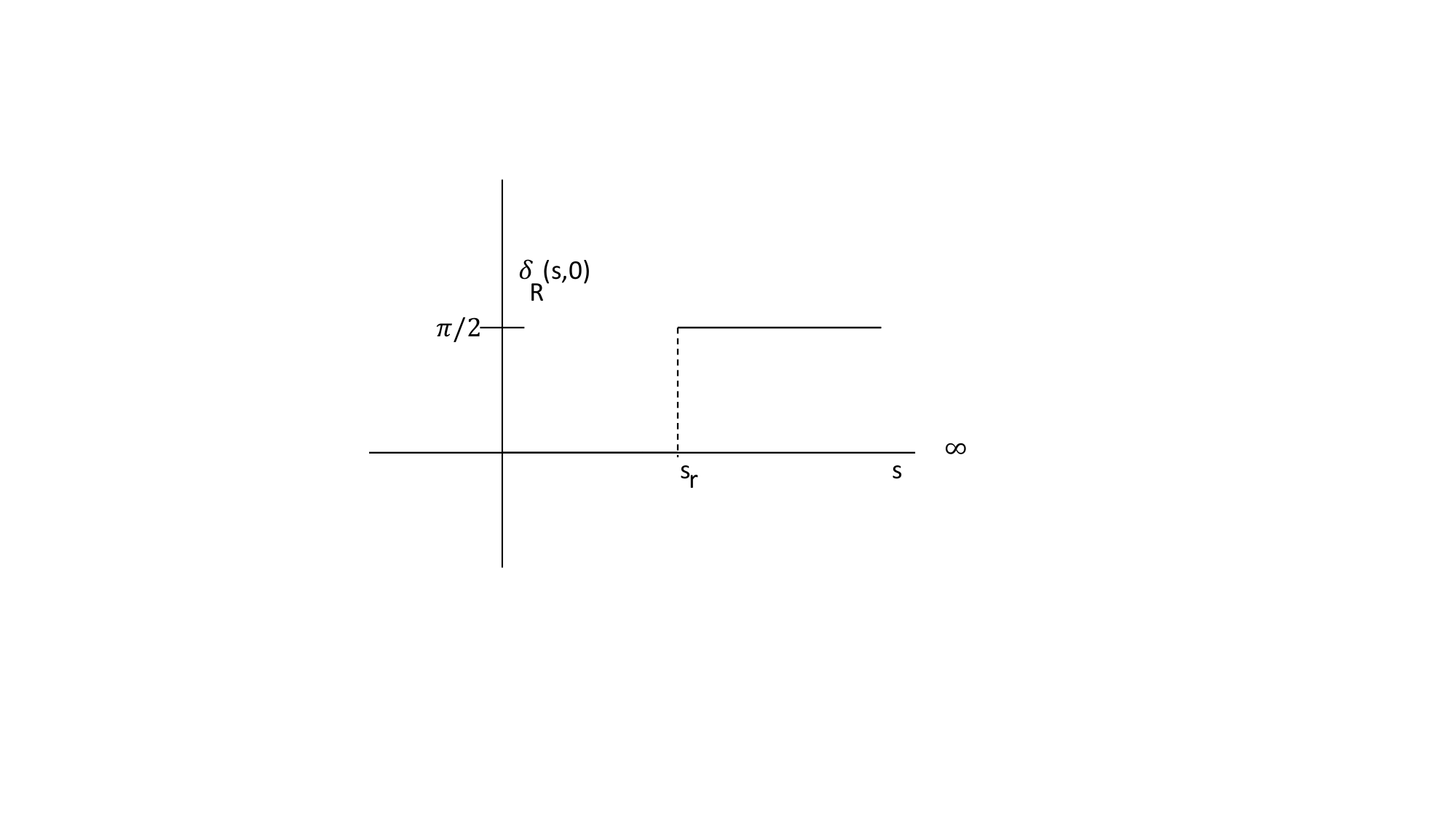}}
	\caption[ch2]{\small Critical behavior of the scattering phase real part $\delta_R(s,0)$ with energy. }
\end{figure}
The respective energy dependence of the function $R(s)$ at the energies beyond the LHC energy region is given by Eq. (\ref{mapr}).  Since the function $g(s)$ increases with energy the reflective ability $R(s)$ becomes nonvanishing and increases also,  $R(s)\to 1$  and $dR(s)/ds\to 0$ at $s\to\infty$. 

\textcolor{blue}{Unitarity equation written  in the impact parameter representation allows one to get the following formula for the reflective ability 
	\begin{equation}\label{rab}
		R(s)=|S(s,0)|=\sqrt{1-4h_{inel}(s,0)},
	\end{equation}
	where $h_{inel}$ stands for the inelastic overlap function.  Eq. (\ref{rab}) is valid without reference to the pure imaginary amplitude. Thus, there is a possibility to extract the reflective ability from the ``experimental values'' of the function $h_{inel}$ \cite{l2}. }
\begin{figure}[hbt]
	\begin{center}
		
		\resizebox{8cm}{!}{\includegraphics*{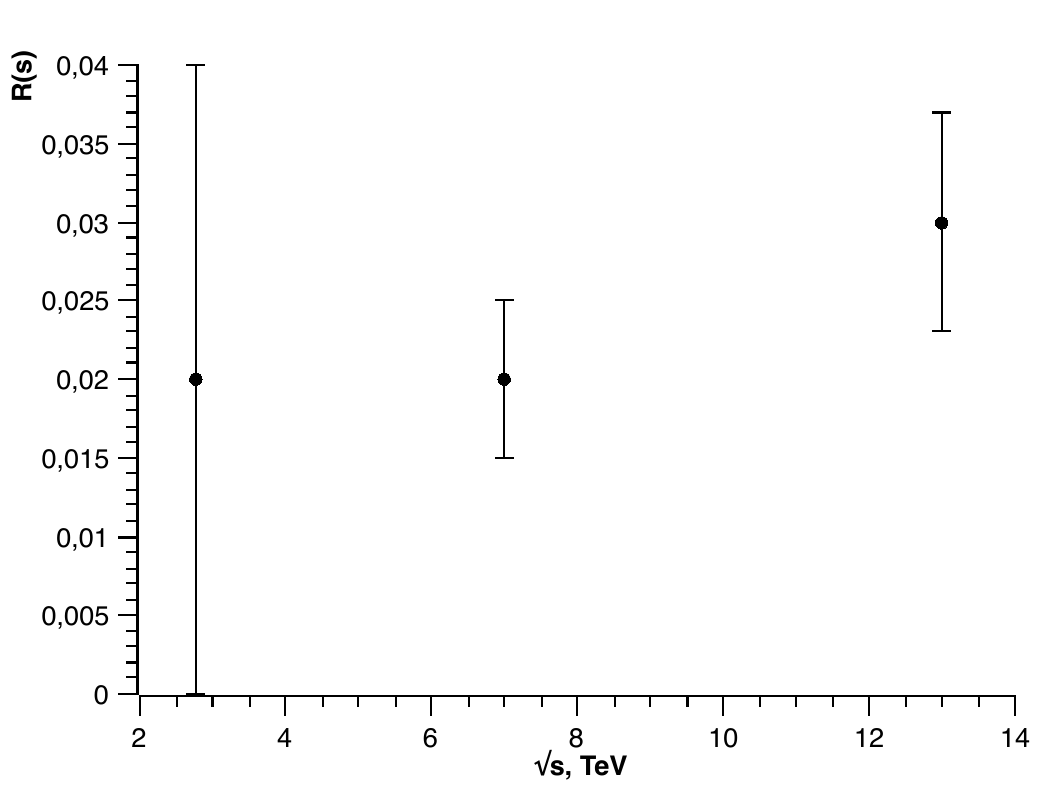}}
	\end{center}
	\hspace{3cm}
	\vspace{-0cm}
	\caption[ch2]{\textcolor{blue}{\small The $R(s)$ values for the three LHC enegries ($\sqrt{s}=2.76,\, 7,\, 13$ TeV)  The  ``experimental'' points are calculated from   $h_{inel}$ values extracted from reference \cite{l2}.}}
\end{figure}
\textcolor{blue}{
	As it follows from Fig. 4,  $R(s)$  becomes positive at the LHC energies due to the reflective scattering mode. Appearance of this mode results in the decrease of the function $h_{inel}(s,0)$ with energy from the value of 1/4 to zero  at $s\to\infty$. The inelastic overlap function $h_{inel}(s,b)$ aquires peripheral dependence on the impact parameter with maximum 1/4 at $b>0$ and it corresponds to a black ring picture in asymptotics.} \textcolor{blue}{The decrease of $h_{inel}(s,0)$ with energy has been considered and plotted in \cite{overdep}.}

The reflective scattering requires  $\delta_R(s,0)=\pi/2$ at $s>s_r$. \textcolor{blue}{It is evident from Eq. (\ref{mapi}), from the exponential representation for $S(s,b)$ and  correspondence of reflective scattering  to $U(s,b=0)>1$.  Note that the model--dependent form of the function $U(s,b)$ is given by Eq. (\ref{usb}).}

Fig. 5 shows critical behavior of $\delta_R(s,0)$.  Dependence of the   phase imaginary part , Eq. (\ref{exp}),  in the reflective region is related to dependence of the reflective ability $R(s)$ of the interation region:
\begin{equation}
	\delta_I(s,0)=-\frac{1}{2}\ln R(s).
\end{equation}
Decreasing energy dependence \textcolor{blue}{ of $R(s)$   sketched also at Fig. 5 } results from increase of the reflective ability
\textcolor{blue}{Respectively, $R(s)$ increases from $R(s)=0$ at $s=s_r$ ($s_r\simeq s_{LHC}$) to the value $R(s)=1$ when $s\to\infty$. It is easy to observe from Eq. (\ref{rab}) and Fig. 4 that $dR(s)/ds>0$ at $s_r\simeq s_{LHC}$. When $s\to\infty$, the approximation 
	\begin{equation}\label{app}
		R(s)\simeq 1-2h_{inel}(s,0) 
	\end{equation}
	can be used. Eqs. (\ref{rab}) and (\ref{app}) highlight the connection of the reflective ability with the asymptotics.
}

Increase of the reflective ability implies  slow down of the mean multiplicity growth at the energies beyond the LHC
\cite{sld} with respective enhancement of the elastic scattering at large transferred momenta (deep--elastic scattering) \cite{cent}. It also positevely correlates  with energy dependence of the ratio of the experimentally observable quantities $Y(s)\equiv \sigma_{tot}(s)/16\pi B(s)$, where 
$\sigma_{tot}(s)$ is the total cross-section  and $B(s)$ is the slope of the differential cross-section of elastic scattering at $-t=0$. Both these quantities are the integrals over impact parameter asymptotically proportional to $\ln^2 s$. The ratio $Y(s)$ was suggested to be interpreted as an effecive interaction intensity (since the ratio effectively eliminates    effect of the interaction radius increase in the total cross--section growth)  and is expected to cross the black disk limit value $1/2$ at $\sqrt{s}\simeq 10^8$ GeV \cite{intens}. Its observed energy increasing dependence at  available energies is just another indication of the asymptotics remotness and is connected to  the  reflective ability increase. 

The connection of reflective ability $R(s)$ with the function  $Y(s)$ turns into an approximate equality in the case of the Fermi-like impact parameter form (flat at small and moderate impact parameters and zero at large ones) of  the elastic profile function. Such form  generates multiple dips and bumps in the respective differential cross--section of nucleons' elastic scattering at very high energies. We can say that this form is common for the elastic scattering of nuclei at current energies \cite{hen}, nuclei elastic scattering at current energies provides a window to hadron elastic scattering at asymptotic energies.
\begin{figure}[hbt]
	\hspace{0cm}
	\vspace{-2cm}
	\resizebox{16cm}{!}{\includegraphics*{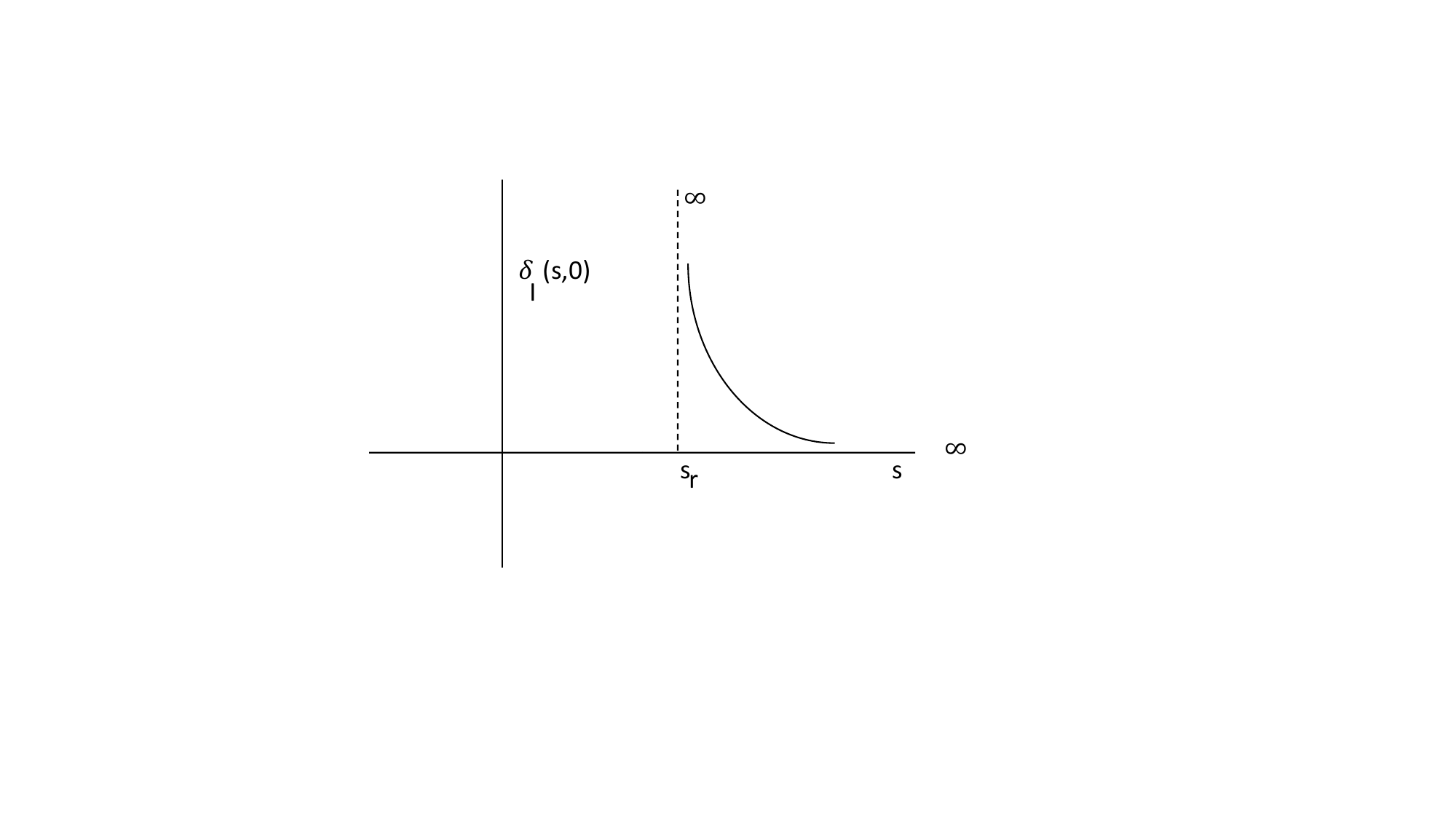}}
	
	\caption[ch2]{\small Energy dependence of the scattering phase imaginary part $\delta_I(s,0)$  in the reflective region at $s>s_r$. }
\end{figure} 
\textcolor{blue}{Qualitative discussion of such behavior of the function $\delta_I$ can be found in \cite{07} while numerical estimates of the value $s_r$ correspond to Fig. 4.}
One can  estimate \textcolor{blue}{also} the energy values when the  reflective ability becomes close to  its asymptotic value -- unity where the relation 
\begin{equation}
	R(s)\simeq Y(s)
\end{equation}
takes place.
It should happen at   the energies around the value of $\sqrt{s}\simeq 10^{10}$ TeV which corresponds to  asymptotic scattering regime when the asymptotical theorems, in particular, $\sigma_{tot}(s)\sim \ln^2 s$ are fulfilled.

Eqs. (\ref{mapi}) and (\ref{usb}) imply that reflective domain of the interaction region enlarges with energy spreading into periphery  of  the impact parameters  with the rate of $\sim\ln s$.
It should be noted that the absorptive scattering mode does not imply the appearance of reflective  domain in  hadron  interactions picture  at all the energies. It is not surprizing since absorption does not cover the whole allowed by unitarity region of the amplitude variation \cite{bla}. 

\textcolor{blue}{A  short comment now on the relation of  $R(s)$ with the ratio of $\sigma_{el}/\sigma_{tot}$. According to Eq. (\ref{app}), the increasing energy dependence of this ratio is connected to emergence of the black ring picture  discussed in \cite{overdep}.}

\section*{Conclusion}
Quantitave analysis of the LHC experimental data provides an evidence for  appearance of the reflective ability ($S<0$) at the energy $\sqrt{s}=13$ TeV \cite{l2} which corresponds to  the energy value $\sqrt{s_r}$.  \textcolor{blue}{This conclusion can be also made from Fig. 4 despite quite large error bars.}

Decrease of the imaginary part $\delta_I(s,0)$  with energy should not be interpreted as an increase of the interaction region transparency (due to deepening of the local minimum of the inelastic overlap function at $b=0$), but instead of it, this deepening means the reflective ability increase which corresponds to repulsion  in hadron interaction dynamics due to  an inner hadron core presence (it, with  reservations, corresponds  to the negative Wigner time delay  $Q$\cite{wig,gol}).

We turn to the interpretation of  the reflective ability increase.   It was proposed \cite{jpg19} to associate  the reflective scattering mode appearance with formation of a color--conducting medium  in the intermediate state of the hadron collision occured at sufficiently high energies and small  impact parameters regarding temperature as depending on the the initial energy and impact parameter of collison. This media is treated as a consisting of the free colored objects.  Therefore, a color--conducting media emerges instead of a color--insulating media  at lower energies.  

Using analogy with scattering of the electromagnetic wave by metals\footnote{Reflective ability of metal is proportional to its electic conductivity due to presence of free electrons.} one can correlate energy increase  of the reflective ability with increase of color conductivity of the deconfined medium. The above analogy is based on similarity of gluons and photons, i.e.  on the replacement of an electromagnetic field of QED by a chromomagnetic field of QCD. 

The  appearance of a color--conducting media  is associated with  the critical dependence of the  phase real part $\delta_R(s,0)$, while {\it increase of the reflective ability} is associated with decrease   of the imaginary part of the scattering phase $\delta_I(s,0)$ at the energies in the region just beyond the LHC.

\end{document}